\documentclass[aps,prb,superscriptaddress,reprint,showpacs,amssymb,amsmath,floatfix,citeautoscript,longbibliography]{revtex4-1}

\usepackage[usenames,dvipsnames]{xcolor}

\usepackage[normalem]{ulem}
 \newcommand{\stkout}[1]{\ifmmode\text{\sout{\ensuremath{#1}}}\else\sout{#1}\fi}

\usepackage{float}
\usepackage{graphicx}% Include figure files
\usepackage{dcolumn}% Align table columns on decimal point
\usepackage{bm}% bold math
\usepackage{multirow}
\usepackage{amsmath,amssymb}

\begin{document}

\title{Resonant x-ray diffraction from chiral electric-polarization structures}

\author{S. W. Lovesey}
       \affiliation{Diamond Light Source, Didcot OX11 0DE, United Kingdom}
       \affiliation{ISIS Facility, RAL, Oxfordshire OX11 0QX, United Kingdom}
   
\author{G.~\surname{van~der~Laan}}
\email[Corresponding author: ]{gerrit.vanderlaan@diamond.ac.uk}
       \affiliation{Diamond Light Source, Didcot OX11 0DE, United Kingdom}

\date{\today}

\begin{abstract}
Heterostructures of PbTiO$_3$/SrTiO$_3$ superlattices have shown the formation of ``polar vortices'', in which a continuous rotation of ferroelectric polarization spontaneously forms. Recently, Shafer {\it{et al.}} [Proc.\ Natl.\ Acad.\ Sci.\ (PNAS) {\bf{115}}, 915 (2018)] reported strong {\it{non-magnetic}} circular dichroism (CD) in resonant soft x-ray diffraction at the Ti $L_3$ edge from such superlattices. The authors ascribe the CD to the chiral rotation of a polar vector. However, a polar vector is invisible to the parity-even electric-dipole transition which governs absorption in the soft x-ray region. A realistic, non-magnetic explanation of the observed effect is found in Templeton-Templeton scattering. Following this route, the origin of the CD in Bragg diffraction is shown by us to be the chiral array of charge quadrupole moments that forms in these heterostructures. While there is no charge quadrupole moment in the spherically symmetric $3d^0$ valence state of Ti$^{4+}$,  the excited state $2p_{3/2}3d(t_{2g})$ at the Ti $L_3$ resonance is known to have a  quadrupole moment. Our expressions for intensities of satellite Bragg spots in resonance-enhanced diffraction of circularly polarized x-rays, including their harmonic content, account for all observations reported by Shafer {\it{et al.}} We predict both intensities of Bragg spots for the second harmonic of a chiral superlattice and circular polarization created from unpolarized x-rays, in order that our successful explanation of existing diffraction data can be further scrutinized through renewed experimental investigations. The  increased understanding of chiral dipole arrangements could open the door to switchable optical polarization.
\end{abstract}

%\pacs{}

\maketitle

%%% {\it{\blue{ --Introduction--  }

\section{ INTRODUCTION}

Ferroelectric thin films and superlattices are currently the subject of intensive research due to their technological applications and  properties that are of fundamental scientific interest.
Ferroelectric materials are characterized by a spontaneous bulk electric polarization, {\bf{P}}, which is switchable by an applied electric field, {\bf{E}}. Typical ferroelectric $P$-$E$ hysteresis loops are akin to  $M$-$H$ hysteresis loops in ferromagnets with  bulk magnetization, {\bf{M}}, and applied field, {\bf{H}}. However, microscopic features that lead to ferromagnetism and ferroelectricity are quite distinct. Ferroelectrics have an asymmetry in the electronic {\it{charge}}, whereas ferromagnets have an asymmetry in electronic {\it{spin}} \cite{Lines2001, Spaldin2003}.

The ground-state structure of most ferroelectrics is due to small atomic displacements from the centrosymmetric paraelectric phase that the structure adopts above the Curie temperature, $T_\mathrm{C}$. In conventional transition-metal ferroelectrics the polar phase is stabilized by lowering the energy of the chemical bond formation, which tends to be favored by empty $d$ orbitals and consequently the absence of magnetism. The tendency of a material to ferroelectric instability can be described by the  pseudo-Jahn-Teller effect \cite{Polinger2017}. The balance between the positive and negative second-order terms usually results in off-centering for $d^0$ cations, such as for Ti$^{4+}$ in the prototypical ferroelectric BaTiO$_3$. Above the ferroelectric   $T_c$  the system is cubic; below $T_c$ it is tetragonal. In transition metals with partially filled $d$ shells, on the other hand, the repulsive Coulomb interactions are stronger than any energy gain from chemical bond formation, and ferroelectric off-centering does not occur.

The largest class of all ferroelectric materials are the perovskites, which are corner sharing oxygen octahedral compounds, with the mineral name of calcium titanate (CaTiO$_3$) having a structure of the type ABO$_3$. Examples are barium titanate (BaTiO$_3$), lead titanate (PTO), lead zirconate titanate (PZT), lead lanthanum zirconate titanate (PLZT), and strontium titanate (STO), all containing Ti $d^0$.

Superlattices of ferroelectric layers offer the possibility to tune the ferroelectric properties while maintaining perfect crystal structure and a coherent strain, even throughout relatively thick samples  \cite{Bousquet2008}. This tuning is achieved in practice by adjusting both the strain, to enhance the polarization, and the composition, to interpolate between the properties of the combined compounds. The balance of elastic, electrostatic, and gradience energies yield a   complex phase diagram \cite{Aguado2011}. In-plane and out-of-plane {\bf{P}} in the different ferroelectric layers leads to an interface that shows polarization rotation. Near the domain walls the local polarization pattern displays a continuous polarization rotation  around the domain walls, connecting two 180$^\circ$ domains \cite{Bousquet2008}. 
  ``Polar vortices'', where the sense of rotation can be reversed by an external electric field, were found in (PbTiO$_3$)$_n$/(SrTiO$_3$)$_n$ superlattices  (often referred to as PTO/STO)  grown on DyScO$_3 (001)$ substrate with repetitions of $n$  unit cells  \cite{Yadav2016,Damodaran2017}.

The soft x-ray wavelength (1-3 nm) is idealy matched to the periodicity of the lateral vortex modulations ($\sim$10 nm).
Recently, Shafer {\it{et al.}}\ \cite{Shafer2018} reported strong non-magnetic circular dichroism (CD) in the resonant x-ray diffraction (RXD) at the Ti $L_3$ edge ($2p \to 3d$ transition) from the chiral electric polarization texture of PTO/STO superlattices with  $n$ = 10-16 unit cells. The authors ascribe the observation of CD to the helical rotation of the electric polarization vector. While the experimental result is not in question, a sound interpretation of the effect is  absent so far. Note that no magnetization is involved to create the CD.  Hence the question that we will address in this paper is the following: What is the optical mechanism of the CD effect  here? Since the underlying physical origin has remained unclear in the paper by Shafer {\it{et al.}} \cite{Shafer2018}, we will present a plausible theoretical interpretation of the phenomenon, which will inspire further experiments on ferroelectric materials. We start by revisiting and indeed excluding some alternative explanations.

Optical transitions can be expanded in multipole terms like electric dipole (E1), electric quadrupole (E2), magnetic dipole (M1), together with higher-order multipole terms that can normally be neglected. \cite{vanderLaan2006} In the soft-ray region, electric-dipole (E1) transitions are by far the most dominant contribution in the light-matter interaction. As a general rule, dichroism can only exist if there is no symmetry that reverses one measurable observable but leaves the rest of the system unchanged.  For instance, shining circularly polarized x-rays onto a magnetic material gives rise to x-ray magnetic circular dichroism (XMCD) in x-ray absorption spectroscopy (XAS) \cite{vanderLaan2014}. The magnetization, {\bf{M}}, is given by an axial vector, which is time-odd and parity-even. The time-reversal operator reverses both the magnetization and the photon helicity. Chiral structures, such as the chiral magnetic domain structure in ultrathin FePd films \cite{Durr1999} and the skyrmion lattice of Cu$_2$OSeO$_3$ \cite{Zhang2016}, show strong CD of the magnetic satellite peaks in RXD. 

Satellite peaks however are no longer expected to be present when the magnetization {\bf{M}} is replaced by the electric polarization {\bf{P}}, which is a polar vector that is time-even and parity-odd. This means that {\bf{P}} is invisible for soft x-rays, since the transition probability is not fully symmetric. The parity-odd {\bf{P}} can only be observed under parity breaking, i.e., interference between even and odd terms in the optical transition. Natural circular dichroism is forbidden for pure electric-dipole transitions (E1-E1) but instead requires optical activity of either E1-M1 or E1-E2 transitions.  M1 transitions can  only be large in the visible and microwave region, but become vanishingly small in the soft and hard x-ray region, due to the restriction imposed by the monopole selection rule for the radial part. 

E1 and E2 transitions from a Ti $2p$ core state are allowed to even $3d$ and odd ($4p$, $4f$) excited states, respectively. Thus the E1-E2 interference requires a substantial parity mixing between the Ti $3d$ and $4p$ states, for which inversion symmetry at the Ti site has to be broken. The E2 transition is only significant for hard x-rays ($>$3 keV). Atomic calculations for Ti$^{4+}$ using Cowan's code \cite{Cowan1981} show that E1-E2 is $\sim$1\% of E1-E1 and can be safely neglected.
A further discussion of  E1-E2 is given in Appendix~\ref{sec:appA}.
 
Having established that the dominant contribution arises from the E1-E1 transition, and dismissed dipoles {\bf{M}} and {\bf{P}} as possible origins, another possibility is given by the charge quadrupole moment, $\langle {\bf{T}}^2 \rangle$. This anisotropic charge distribution gives rise to x-ray linear dichroism (XLD) in XAS, and Templeton-Templeton (T\&T) scattering in RXD. 

\begin{figure}[t]
   \centering
%%%   trim: left, bottom, right, top  %%%
\includegraphics[trim = 20mm 10mm 20mm 1
0mm,clip, width=75mm, angle=0]{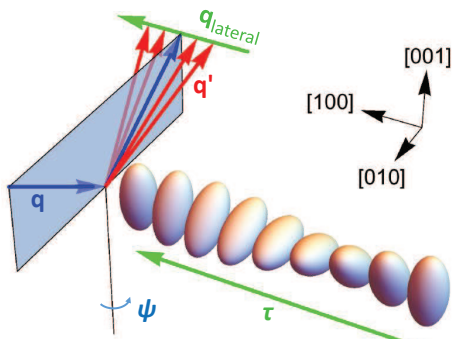} 
   \caption{
Self-organized arrays of electric polarization textures in (PbTiO$_3$)$_n$/(SrTiO$_3$)$_n$ superlattices exhibit a chiral RXD pattern. The chiral array of quadrupole moments produce diffraction satellites that decorate the specular reflection along the lateral direction, [100], for x-rays tuned at the Ti $L_3$ resonance. The specular scattering plane is shown in light blue. Blue arrows give the primary and secondary beam. Red arrows give the satellite peaks, which show non-magnetic CD. $\tau$ represents the helix wavevector $2 \pi/d$.
The sample is rotated about the wavevector  $\mathbf{q} - \mathbf{q}'$  by an angle $\psi$.
The physical picture  differs from that by Shafer {\it{et al.}} \cite{Shafer2018}, who uses polar vectors, representing {\bf{P}}, instead of quadrupole moments, representing the charge anisotropy.
    }
   \label{fig:1}
\end{figure}

There will be no circular dichroism from $\langle {\bf{T}}^2\rangle$ at a single site, since this tensor  is time-even. It is easy to see that $\langle {\bf{T}}^2\rangle$ would have to contain an imaginary part, which   for a centrosymmetric site is unphysical. 
However,   an ordered chiral configuration of quadrupole moments    contains an imaginary component. The spherical tensor $\langle {\bf{T}}_Q^2\rangle$, with projections $Q$ = 0, $\pm$1, $\pm$2, transforms under space rotation like a spherical harmonic $Y^2_Q(\theta,\varphi)$ with azimuthal $\varphi$ dependence $\exp ({\mathrm{i}}Q \varphi)$. An angular rotation by  $2 \pi/n$ of the spherical tensor gives a phase shift $\exp (2 \pi {\mathrm{i}} Q/n)$ and a translation over {\bf{R}} gives a phase factor $\exp ({\mathrm{i}}{\bf{\tau}} \cdot {\bf{R}})$.  Properly taking  all phase factors  into account in the coherent sum for the diffracted intensity from a chiral structure of quadrupole moments,  incident x-rays of left- and right-circularly polarization with phase factors $\exp(\pm \mathrm \pi/2)$ give different intensities. Thus, there will be  non-magnetic CD in RXD.  The scattering geometry of the measurement is schematically illustrated in Fig.~\ref{fig:1}, showing a helical array of  quadrupoles with wavevector $\tau$. The chiral satellites appear in the lateral direction  with respect to the scattering plane.

 By tuning the x-ray energy near the Ti $L_{2,3}$ resonance ($2p \to 3d$ transition), the RXD becomes sensitive to he anisotropic electronic structure of the distorted TiO$_6$ octahedra. Previous  XAS studies \cite{Arenholz2010} of ferroelectric Pb(Zr$_{0.2}$Ti$_{0.8}$)O$_3$  already showed a large XLD, evidencing  charge anisotropy of the Ti site, despite the $d^0$ ground state (see Sec.~\ref{sec:quadrupole}).

In the following, we will present a precise analysis of the CD effect in RXD and make some further predictions. For easy of reference, we will consistently use the notation laid out in the review of Lovesey {\it{et al.}} \cite{Lovesey2005}

%%%%%%  BEGIN FIGURE 1 %%%%%
%%%%%%  END FIGURE 1 %%%%%

\section{RESONANCE-ENHANCED DIFFRACTION}

\subsection{Preamble and Survey}

An approximation of spherical ions in a material is central to electronic structure determinations from Bragg diffraction patterns. Departures from spherical symmetry in atomic charge distributions are announced by the addition of weak Bragg spots. For example, Adachi {\it{et al.}} \cite{Adachi2002} report basis-forbidden reflections in Thompson scattering from dysprosium borocarbide. A helpful enhancement of weak spots can be captured by tuning the primary x-ray beam to an atomic resonance. Moreover, resonance-enhanced diffraction can rotate primary polarization, and thus depend on photon helicity, whereas this is not possible in Thomson scattering that is diagonal in polarization states. Absence of translation symmetry also generates basis-forbidden reflections, or satellite reflections. Bragg spots for a composite material that we examine are consequences of departures from spherical symmetry in electronic structure and, also, the absence of translation symmetry in a chiral structure. Oscillators make up the energy profile we employ, leaving us the task of calculating structure factors for diffraction. A division of the energy profile from atomic properties of valence electrons in their ground state is an approximation that is consistent with the fast-collision concept and angular isotropy of the core state \cite{Hannon1988, Lovesey1997, Lovesey2005}.
	
A theory of resonance-enhanced diffraction is set out by Dmitrienko \cite{ Dmitrienko1983, Dmitrienko1984}, and Templeton \& Templeton \cite{Templeton1985a, Templeton1985b, Templeton1986} report the first relevant data, e.g., tetragonal K$_2$PtCl$_4$ ($P4/mmm$-type) and sodium bromate ($P2_13$-type). Formulations of RXD found in the cited papers use classical optics and physical properties of crystals with Cartesian tensors, as in the treatise by Nye \cite{Nye1985}, and no attempt is made to calculate an energy profile. An atomic theory appeared shortly afterwards \cite{Hannon1988} in response to the publication of experimental data for RXD by an incommensurate magnetic motif. Hannon {\it{et al.}} \cite{Hannon1988,Luo1993} show that resonance-enhanced diffraction provides sensitivity to a wealth of electronic properties of a material, and their insight is the basis of current atomic theories. 

For the sake of demonstration, Hannon {\it{et al.}} \cite{Hannon1988} tackle the formidable task of describing electronic structure in the scattering length by imposing cylindrical symmetry at sites occupied by resonant ions. In consequence, the task is reduced to a vector model using a single material-dipole for the resonant ion. This early approximation to the scattering amplitude is not universal and it is inadequate for the interpretation in hand. Reviews of many applications of resonance-enhanced diffraction include Refs.~\onlinecite{Carra1994,Lovesey2005, Dmitrienko2005, vanderLaan2008,Collins2010,Matsumura2013}

As we have mentioned, a vector model that imposes cylindrical symmetry on electronic charge distributions is too restrictive for present purposes. Instead, we place no restrictions on local charge distributions associated with resonant Ti ions in the superlattices. A coherent sum of unrestricted distributions to form of a helix is shown to account for pivotal properties of the diffraction pattern observed with circularly polarized x-rays. Assuming intensities of Bragg spots are enhanced by an E1-E1 absorption event, it follows that Ti charge distributions engaged in diffraction must be electronic quadrupoles. Our appeal to T\&T scattering for an interpretation of the diffraction pattern produced by PTO/STO superlattices illuminated with circularly polarized photons mirrors previous work for crystalline materials \cite{Tanaka2010, Lovesey2013}.  

\subsection{Charge-quadrupole moment for Ti$^{4+}$}
\label{sec:quadrupole}

We shall first explain that the Ti$^{4+}$ ground state $3d^0$ is not a limitation for having access to a quadrupole moment. After all,  Ti $L_{2,3}$ XAS of lead zirconate titanate   shows linear dichroism  \cite{Arenholz2010}.
The measured Ti $L_{2,3}$ XAS exhibits the characteristic structure of a Ti$^{4+}$ $3d^0$ configuration due to E1 transitions to $2p^53d^1$ final states \cite{vanderLaan1992}. The $2p$ spin-orbit interaction splits the spectrum into $2p_{3/2}$ ($L_3$) and $2p_{1/2}$ ($L_2$) peaks with energy separation of $\sim$5.5 eV. These peaks are further split by crystal-field interaction, i.e., the electrostatic potential, $V$, due to the neighboring lattice sites acting on the $3d$ orbitals. In octahedral ($O$) site symmetry the $e_g$ orbitals of Ti point toward the oxygen ligands, while the $t_{2g}$ orbitals point in between them, resulting in a lower energy for the latter. With increasing photon energy the four main peaks can therefore be labelled according to their main character as $2p_{3/2}3d(t_{2g})$, $2p_{3/2}3d(e_g)$, $2p_{1/2}3d(t_{2g})$, and $2p_{1/2}3d(e_g)$, cf.\ Fig.~1 in Arenholz {\it{et al.}} \cite{Arenholz2010}  Lowering to tetragonal symmetry splits the $t_{2g}(O)$ states into $b_2=d(xy)$ and $e=d(xz, yz)$ and the $e_g(O)$ states into $b_1=d(x^2 - y^2)$ and $a_1=d(z^2)$. 
This results in a difference between XAS spectra taken with polarization along the $z$ and   ($x$, $y$) directions. 
The multiplet structure contains peaks with mixed character of these crystal-field states \cite{vanderLaan2006}.
The  nominal $2p_{3/2}3d(t_{2g})$ excited state has the  largest quadrupole moment, so that this  resonance is used to tune the photon energy in RXD.
The superlattice is expected to have a continuous tilt of the Ti $t_{2g}$ orbitals relative to polarized x-ray beam.  

\subsection{Scattering length}
	
\begin{figure}[t]
   \centering
%%%   trim: left, bottom, right, top  %%%
\includegraphics[trim = 25mm 0mm 20mm 5mm,clip, width=80mm, angle=0]{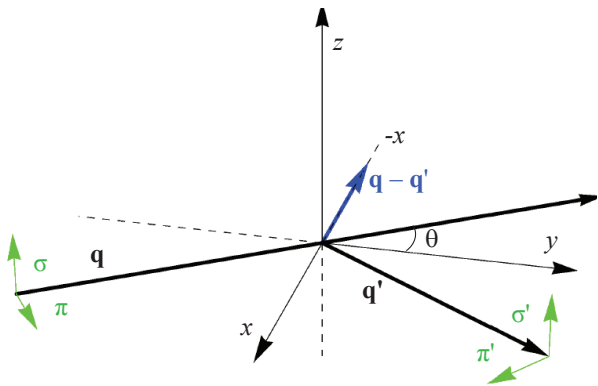} 
   \caption{
  Primary ($\sigma$, $\pi$) and secondary ($\sigma'$, $\pi'$)  states of polarization. Corresponding wavevectors ${\mathbf{q}}$ and ${\mathbf{q}}'$ subtend an angle $2 \theta$. Helicity in the primary beam is proportional to a Stokes parameter $P_2$ that is a time-even pseudo-scalar \cite{Lovesey2005}. Cell edges of the sample and depicted Cartesian coordinates ($x$, $y$, $z$) coincide in the nominal setting, and the sample is rotated to align ${\mathbf{q}}  - {\mathbf{q}}'$ = [001], which is along  $-x$-direction. The desired alignment is rotation by 90$^\circ$ about the $y$-axis, and the sample axis [010] lies in the plane of scattering for the azimuthal angle $\psi$ = 0$^\circ$. 
 }
   \label{fig:2}
\end{figure}

The Kramers-Heisenberg dispersion formula yields a scattering length
\begin{equation}
g = - \frac{r_e}{m} \frac{F_{\mu'\nu}}{(E - \Delta + \frac{1}{2} \mathrm{i} \Gamma)} ,
\end{equation}
\noindent
with the primary x-ray energy $E = 2\pi c \hbar / \lambda$  in the vicinity of an atomic resonance $\Delta$ that has a lifetime $\propto$  $\hbar/ \Gamma$ ($r_e \approx 0.282 \times  10^{-12}$ cm, $mc^2$ $\approx$ 511 keV, and $\lambda$[\AA] $\approx$ $12.4/E$[keV]. $F_{\mu'\nu}$   is an amplitude for polarizations $\mu'$ (secondary) and $\nu$ (primary). Following the usual conventions for polarization states depicted in Fig.~\ref{fig:2}, where $\sigma$ ($\pi$) labels polarization normal (parallel) to the scattering plane. The illuminated sample is rotated about the wavevector $\mathbf{q} - \mathbf{q}'$ by an angle $\psi$, and $\mathbf{q} \cdot \mathbf{q}'$ = $(2 \pi/ \lambda)^2 \cos2 \theta$.

%%%%%%  BEGIN FIGURE 2 %%%%%
%%%%%%  END FIGURE 2 %%%%%

Intensities ${\cal{J}}$ of   Bragg spots proportional to the primary photon helicity (Stokes parameter $P_2$) are derived from
\begin{equation}
{\cal{J}} = P_2 \, \mathrm{Im} \left[F_{\sigma' \pi}^\ast F_{\sigma' \sigma}  + F_{\pi' \pi}^\ast F_{\pi' \sigma} \right] .
\end{equation}
Evidently, the true scalar ${\cal{J}}$ can be different from zero when the combination of scattering amplitudes is both time-even and a pseudoscalar to match discrete symmetries of $P_2$. For the case in hand, $F_{\mu'\nu}$ is a (dimensionless) linear combination of axial quadrupoles $\langle \mathbf{T}^2_Q \rangle$ engaged in diffraction enhanced by an E1-E1 resonance. Expressions we need for $F_{\mu'\nu}$(E1-E1) have been listed by Scagnoli and Lovesey \cite{Scagnoli2009}.

\subsection{Atomic states}

A titanium ion (Ti$^{4+}$) nominally possesses an empty $3d$ shell that accepts a photo-ejected $2p$ electron. For such a configuration of electrons the T\&T scattering is zero, because the $3d$-shell is spherically symmetric with no angular anisotropy. A non-zero result for bulk lead titanate \cite{Arenholz2010} is attributed to the influence of a local electrostatic potential experienced by a Ti ion that splits $3d$ states in levels of lower than octahedral symmetry, as discussed in Sec.~\ref{sec:quadrupole}.

Dependence of $\langle {\mathbf{T}}^2_Q \rangle$ on the total angular momentum of the core state $\overline{J}$ gives rise to sum rules for dichroic signals \cite{vanderLaan1998}. The reduced matrix-element (RME) of the spherical tensor operator is derived in Ref.~\onlinecite{Lovesey1997} and  expressed in Eq.~(73) of Ref.~\onlinecite{Lovesey2005}. For $L$ edges the RME has one component for $\overline{J}$  = 1/2 and three components for $\overline{J}$ = 3/2. Differences in the RMEs can lead to striking differences in intensities of diffraction spots, with intensity at $L_2$ small and beyond observation while $L_3$ intensity is significant. 

The sum rule for quadrupoles that produce T\&T scattering at $L$ edges is obtained by making a sum on $\overline{J}$ in the RME [Eq.~(80) in Ref.~\onlinecite{Lovesey2005}]. One finds
\begin{equation}
\langle {\mathbf{T}}^2 \rangle_{L_3} + \langle {\mathbf{T}}^2 \rangle_{L_2} =
\langle \{ {\mathbf{L}} \otimes {\mathbf{L}} \} \rangle^2/60 .
\end{equation}
Here, $ \{ {{\mathbf{L}} \otimes {\mathbf{L}} }  \}^2$ is a standard tensor product of rank 2 formed with angular momentum ${\mathbf{L}}$, with a diagonal component 
$\{ {{\mathbf{L}} \otimes {\mathbf{L}} } \}^2_0 = [3(L_z)^2 - L(L + 1)]/ \sqrt 6$.
 The sum rule is nothing more than an identity satisfied by the relevant RMEs, and it can be applied to any ground-state wavefunction for the expectation value denoted by $\langle \cdots \rangle$.

\subsection{Bulk lead titanate}

	Displacement of a Ti ion from the centre of the $O$-octahedron below a temperature  $\sim$760 K in bulk PbTiO$_3$ creates a ferroelectric moment. In the polar, acentric tetragonal structure P4mm (\#99) cell lengths $c/a \approx 1.063$, with titanium ions at sites 1a that have symmetry 4mm ($C_{4v}$). The tetrad axis of rotation symmetry renders all electronic quadrupoles diagonal ($Q = 0$).
	
\section{SUPERLATTICES}

\subsection{Diffraction intensities}

Titanium quadrupoles in a superlattice are likely devoid of any local angular symmetry and all five components of $\langle {\mathbf{T}}^2_Q \rangle$ are therefore permitted. Our minimal model uses a helix that propagates along the unit-cell axis [100]. In diffraction the axis is arranged normal to the plane of scattering at the origin of an azimuthal-angle scan. A coherent sum of quadrupoles in such a helix is $C_Q^2(f)$ [in the following denoted as $C_Q(f)$] where integer $f$ is the order of the harmonic, and projections are $Q$ = 0, $\pm$1, $\pm$2. $C_Q(f)$ is constructed with a unit twist in the $y$-$z$ plane $\varphi  = \pm 2 \pi/(2n + 1)$ and the total twist through $(2n + 1)$ units = $\pm 2 \pi n$. A unit wavevector = $2 \pi/d$ where $d$ is the length of a helix. 

An expression for $C_Q(f)$ provided in Ref.~\onlinecite{Scagnoli2009} can be written as
\begin{align}
& (2n + 1) \left[C_Q \pm  C_{-Q} \right] = \left[ \langle {\mathbf{T}}^2_Q \rangle  \pm \langle {\mathbf{T}}^2_{-Q}  \rangle \right] \nonumber \\
& +2 \sum_{m,p} d^2_{Q, p}(m \varphi) \cos [ m \varphi f + \frac{\pi}{2}(Q - p) ]
\left[  \langle {\mathbf{T}}^2_p \rangle  \pm \langle {\mathbf{T}}^2_{-p} \rangle \right],
\end{align}
in which $ d^2_{Q, p}(m \varphi)$ is a purely real element of the standard Wigner rotation matrix, the sum on integer $m$ is in the range 1 to $n$, and projections $p$ = 0, $\pm$1, $\pm$2. $C_Q(f)$ is a complex quantity in the general case and not Hermitian, unlike the quadrupoles $\langle {\mathbf{T}}^2_Q \rangle$ from which it is constructed, i.e., the complex conjugate satisfies $\langle {\mathbf{T}}^2_{-Q} \rangle = (-1)^Q \langle {\mathbf{T}}^2_Q \rangle^\ast$.

	We consider the first and second harmonics with $f$ = 1 and 2, respectively. One finds that $C_Q(f)$ is determined by one of its five components for a given $f$, and these are taken to be $C_1(1)$ and $C_0(2)$ with
\begin{subequations}
\begin{align}
& C_1(1) =  \frac{1}{ \sqrt 6} \left[\mathrm{i} \langle \mathbf{T}^2_{xy} \rangle - \langle \mathbf{T}^2_{zx} \rangle\right] ,    \\
& C_0(2) =  \frac{3}{8} \langle \mathbf{T}^2_{zz} \rangle   + \frac{1}{8} \langle \mathbf{T}^2_{xx} - \mathbf{T}^2_{yy} \rangle - \frac{\mathrm{i} }{2} \langle \mathbf{T}^2_{yz} \rangle .
\end{align}
\end{subequations}
In these results, we use Cartesian forms of the quadrupoles $\langle \mathbf{T}^2_{\alpha \beta} \rangle $ that are purely real, with  $\langle \mathbf{T}^2_{zz} \rangle = \langle \mathbf{T}^2_0 \rangle$. Additional quadrupoles epitomize the angular anisotropy of a TiO$_6$ complex participating in a helix. 

Intensities are
\begin{subequations}
\begin{align}
  {\cal{J}}(f \! = \! 1) = & -  P_2 | C_1(1) | ^2 \beta \gamma \cos^3 \theta \cos \psi \sin 2 \psi ,  
\label{eq:6a}  \\
{\cal{J}}(f \! = \!2) = & - \frac{1}{3} P_2 |C_0(2)|^2 \beta \cos \theta \sin \psi  \nonumber  \\
& \times [\alpha (2 + \cos^2 \theta )  - \gamma \cos^2 \theta \cos 2 \psi] .
\label{eq:6b} 
\end{align}
\end{subequations}
Quantities $\alpha$ ($Q$ = 0), $\beta$ ($Q$ = 1), and $\gamma$ ($Q$ = 2) are weights of $C_Q(f)$ that make up a helix, and they are taken to be purely real. Note that $C_0(1)$ = 0 which accounts for the absence of $\alpha$ in ${\cal{J}}(f \! = \! 1)$. The handedness of the helix allows ${\cal{J}}(f)$ to be different from zero chirality in the sample and impressed helicity in the photon beam engage in diffraction. Specifically, intensity arises from interference between odd projections ($Q = \pm1$) and even projections ($Q$ = 0 and $\pm$2) of $C_Q(f)$ as might be expected. 	

Intensities are zero at the origin of an azimuthal-angle scan $\psi$ = 0$^\circ$ where the plane of rotation of the helix and the plane of scattering are parallel. And they are zero for $\psi$ = 180$^\circ$  and maximum for $\psi$ = 90$^\circ$, all of which accords with the observations. A non-zero ${\cal{J}}(f)$ emerges when a projection of the photon wavevector = $2 \pi / \lambda$ matches the helix wavevector = $2 \pi/d$.  More precisely, diffraction occurs for a Bragg condition estimated to be $\sin \psi = \pm \lambda f /(d \cos \theta)$. A change in sign of $\psi$ is equivalent to swapping between helices with opposing twists, which amounts to the statement that the product $(\varphi \sin \psi)$ has one sign. The two properties mentioned agree with reported observations of   ${\cal{J}}(2)$. Intensities are independent of the sign of the angle $\theta$ set by the superlattice.

\subsection{ Diffracted beam polarization}

Mindful of renewed experiments on PTO/STO superlattices, we provide our prediction for polarization in the diffracted beam \cite{Lovesey2005}. A telling probe of chirality within the superlattice is circular polarization, $P'_2$, created from an unpolarized beam. One finds
\begin{align}
& P'_2(f \! = \! 1)_{\mathrm{unpol}}  \nonumber  \\
& \ \ \ \ = \frac{ 4(\gamma / \beta) \cos^3 \theta \cos^2 \psi \sin \psi }
{2 \cos^2 \theta \cos^2 \psi + (\gamma / \beta)^2 (\cos^4 \theta \sin^2  2 \psi + 2 \sin^2 \theta) } ,
\end{align}
\noindent
which shows that $\gamma / \beta$ can be inferred from experimental data. Note that ${\cal{J}}(f \! = \! 1)$ is proportional to the numerator of $P'_2(f \! = \! 1)$. More generally, the primary beam carries linear polarization, $P_3$, and circular polarization, $P_2$. Circular polarization in the diffracted beam satisfies
\begin{align}
P'_2(f \! = \! 1)  &   = \frac{ |C_1(1)|^2 }{ {\cal{J}}_{\mathrm{o}} (f \! = \! 1)}  \nonumber \\
 \times & \left\{ 2 \beta \gamma \cos \theta \cos^2 \psi \sin \psi [ \cos^2 \theta + P_3 (\sin^2 \theta + 1) ]  \nonumber  \right. \\
& \  \ + \left. P_2 (\gamma^2 \sin^2 \theta - \beta^2 \cos^2 \theta \cos^2 \psi) \right\} ,
\end{align}
\noindent
where ${\cal{J}}_{\mathrm{o}} (f \! = \! 1)$ is the total intensity. 
If the primary beam possesses linear polarization alone $(P_2 = 0)$, 
\begin{align}
{\cal{J}}_{\mathrm{o}} (f \! = \! 1) & = \frac{1}{2} | C_1(1)|^2   \nonumber \\
 \times  & \left\{ 2\beta^2 \cos^2 \theta \cos^2 \psi  + 2 \gamma^2 \sin^2 \theta \cos^2 2 \psi  \right.\nonumber \\
 + & \left.    \gamma ^2 \sin^2 2 \psi [1 + \sin^4 \theta + P_3 \cos^2 \theta (\sin^2 \theta + 1)] \right\} .
\end{align}
\noindent
A Stokes parameter $P_3 = +1$ corresponds to 100\% linear polarization normal to the plane of scattering obtained with x-ray synchrotron sources, to a good approximation.

Corresponding expressions for the second harmonic are very complicated, as might be anticipated from the foregoing lengthy results for $f $ = 1.

\section{DISCUSSION and Conclusion}

We have presented a highly plausible explanation for the appearance of non-magnetic circular dichroism (CD) in resonant x-ray diffraction (RXD) of PbTiO$_3$/SrTiO$_3$ superlattices, which show the formation of ``polar vortices''. The diffraction effect is due to the chiral array of charge quadrupole moments that forms in these heterostructure, which contribute Templeton-Templeton scattering of circularly polarized x-rays. The necessary condition for a quadrupole moment in the resonant state is fulfilled, as previously demonstrated by x-ray linear dichroism (XLD). Additional tests of our explanation might include studies of scattering by the second harmonic of the chiral structure, and a measurement of polarization created from an unpolarized x-ray beam.

 While usually magnetism is at the origin of  circular dichroism in RXD, our analysis  shows that this is not the only possible source. Chiral charge distributions, such as found sometimes at interfaces and surfaces, can also lead to CD in RXD, where the intensity due to the E1-E1 transition can easily exceed that of the E1-E2 interference. Therefore, our analysis will be applicable to a wider class of materials.

A useful analogy for our achievement is to liken it to interpreting diffraction by a molecular crystal. We make calculations for the correct molecular unit that we submit is a helix of quadrupoles. But we are unable to assign a refined global structure for the units in the superlattice with severely limited diffraction data at this time.  
In the case of a  skyrmion lattice, computer simulations of the coherent diffraction of all the magnetic moments, taking into account their positions and directions in the unit super cell with respect to the x-ray beam, has been very successful  \cite{Zhang2018}. A similar approach---but replacing the vector {\bf{M}} by the quadruple moment $\langle {\mathbf{T}}^2 \rangle$  in the coherent sum---can be used to simulate the electric-polarization lattice.

Our model, which previously has only been applied to crystalline structures, has been extended for heterostructures, which are usually incommensurate with the underlying lattice periodicity. A future proposed step is to measure these effects on multiferroics, where normally the central cation has partially filled $d$ states. Contributions from the quadrupole moment can be separated from those from the magnetization by measuring the angular dependence and the polarization of the diffracted beam. 

Multiferroics, which have simultaneous magnetic and ferroelectric ordering, contain all the potential applications and basic scientific interest of their parent ferroelectric and ferromagnetic materials, as well as a whole range of new phenomena and potential technologies resulting from interactions between the two orderings. The functionalization of multiferroics leads to spintronics applications. \cite{Bea2008} From a practical standpoint, if existing magnetism technologies could be tuned or controlled with electric instead of magnetic fields, large improvements in miniaturization and power consumption should be expected.

\begin{acknowledgments}
 
SWL is grateful for correspondence with Dr K S Knight on the structure of lead titanate. Professor S P Collins and Dr V Scagnoli advised on applications of specular reflection. GvdL acknowledges Dr P Shafer and Professor E Arenholz for a discussion of the measurements.
\end{acknowledgments}

\appendix
\section{E1-E2 transition}
\label{sec:appA}

Here we provide further details conserning the E1-E2 transition, which is also allowed but gives only a small contribution.

The ratio of E1-E2 to E1-E1 amplitudes mentioned in the main text is
\begin{equation}
{\cal{R}} = \frac{\alpha E}{2 R_\infty} \frac{\langle 2p|R^2|4p \rangle }{ \langle 2p|R^1|3d \rangle},	
\label{eq:A1}
\end{equation}
\noindent
where $\alpha$ and $ R_\infty$ are the fine-structure constant and the Rydberg unit of energy, respectively, and $E \approx 457$ eV at the Ti $L_3$ edge. The two radial integrals are measured in units of the Bohr radius $a_0$, so their ratio in Eq.~(\ref{eq:A1}) is dimensionless. In so far as hydrogenic forms of radial wavefunctions are appropriate for the photo-ejected $2p$ electron and empty $3d$ and $4p$ valence states $\langle 2p|R^2|4p \rangle / \langle 2p|R^1|3d \rangle$ = $-1.66/Z_{\mathrm{o}}$, where $Z_{\mathrm{o}}$ is the effective core charge seen by the jumping electron. Taking $Z_{\mathrm{o}}$ = 18, appropriate for the argon core of Ti$^{4+}$, yields an estimate $\langle 2p|R^2|4p \rangle / \langle 2p|R^1|3d \rangle$ = $-$0.092, which is in line with atomic calculations.
Indeed, Cowan's atomic Hartree-Fock code with relativistic corrections \cite{Cowan1981} gives the radial integrals for Ti$^{4+}$   as $\langle 2p|R^1|3d\rangle$ = $-$0.26266 $a_0$,  $\langle 2p|R^2|4p\rangle$ = 0.02613 $(a_0)^2$ [and for completeness we also give $\langle 2p|R^2|4f\rangle$ = 0.01208 $(a_0)^2$], which amounts to ${\cal{R}}$ $\approx$ $-$0.012, so that E1-E2 is about 1\% of E1-E1.  

Diffraction enhanced by an E1-E2 event accesses time-even polar multipoles 
$\langle {\mathbf{U}}^K_Q \rangle$   with ranks $K$ = 1, 2, and 3. Polar multipoles satisfy sum rules akin to their parity-even counterparts and full results can be found in Refs.~\onlinecite{Lovesey2010,Lovesey2012}.
The dipole $\langle {\mathbf{U}}^1 \rangle$  is related to the atomic displacement of the resonant ion while the quadrupole $\langle {\mathbf{U}}^2 \rangle$ is more complicated and fully understood. The first harmonic $f  \! = \! 1$ contains all allowed multipoles whereas the dipole is absent in the second harmonic $f \! = \! 2$, as expected. By way of orientation to E1-E2 enhanced diffraction by a helix we report intensities calculated with leading dipoles for a given harmonic. The analogue of Eq.~(\ref{eq:6a}) is the estimate 
\begin{align}
 {\cal{J}}( {\mbox{E1-E2}}; f \!= \!1)  
 \approx & - \frac{3}{200} P_2  \left[ \langle {\mathbf{U}}^1_z \rangle^2 + \langle  {\mathbf{U}}^1_y \rangle ^2 \right]   \nonumber \\
&  \times {\cal{R}}^2 \alpha \beta \sin \theta \sin 4 \theta \sin \psi.
\label{eq:A2}
 \end{align}
While retaining dipoles alone as the leading contribution to the diffraction amplitude the coefficient $\gamma$, the weight of projections $Q = \pm 2$, is understandably absent. Additional comments are given following the companion intensity for $f  \! = \! 2$.

We set aside the polar octupole for the second harmonic and arrive at the estimate
\begin{align}
 {\cal{J}} & ( {\mbox{E1-E2}}; f \!= \!2)    \nonumber \\
& \approx - \frac{1}{3} P_2  \left\{ \frac{1}{16} \left[ 3 \langle {\mathbf{U}}^2_{zz} \rangle + \langle {\mathbf{U}}^2_{xx} - {\mathbf{U}}^2_{yy} \rangle  \right]^2 + \langle {\mathbf{U}}^2_{yz} \rangle^2  \right\}   \nonumber \\
& \  \ \ \ \   \times {\cal{R}}^2
\beta \gamma \cos \theta \sin^2 2 \theta \cos \psi \sin 2 \psi.  
\label{eq:A3}
\end{align}

The contribution to ${\cal{J}}( {\mbox{E1-E2}}; f \!= \!2)$ from quadrupoles proportional $\alpha$, i.e., projection $Q = 0$, adds up to zero. Note that intensities in Eqs.~(\ref{eq:6a}) and (\ref{eq:6b}) for diffraction enhanced by an E1-E1 event are likewise proportional to $\sin \psi$. Hence, all our calculations show that intensity proportional to circular polarization in the primary beam, $P_2$, tracks the lateral wavevector in Fig.~\ref{fig:1} as the superlattice is rotated about the Bragg wavevector. The factor ${\cal{R}}^2$ in Eqs.~(\ref{eq:A2}) and (\ref{eq:A3}) serve to remind that E1-E2 intensities are a very small fraction of the E1-E1 intensities reported in the main text.

%%%%{\it{\blue{Coloured text}}}

% \bibliography{Biblio_SWL}
%merlin.mbs 2010-03-15 4.21a (PWD, AO, DPC)

\end{document}